\documentclass[aps,epsfig,showpacs,twocolumn]{revtex4}
\usepackage{amssymb}
\usepackage{graphicx}

\begin{document}

\title{Scaling Behavior and Variable Hopping Conductivity in the Quantum
Hall Plateau Transition}
\author{Tao Tu}
\email{tutao@ustc.edu.cn}
\author{Yong-Jie Zhao}
\author{Guo-Ping Guo}
\email{gpguo@ustc.edu.cn}
\author{Xiao-Jie Hao}
\author{Guang-Can Guo}
\affiliation{ Key Laboratory of Quantum Information, University of
Science and Technology of China, Chinese Academy of Sciences, Hefei
230026, P. R. China}
\date{\today}

\begin{abstract}
We have measured the temperature dependence of the longitudinal resistivity $%
\rho _{xx}$ of a two-dimensional electron system in the regime of the
quantum Hall plateau transition. We extracted the quantitative form of
scaling function for $\rho _{xx}$ and compared it with the results of
ordinary scaling theory and variable range hopping based theory. We find
that the two alternative theoretically proposed scaling functions are valid
in different regions.
\end{abstract}

 \maketitle

%%%%%%%%%%%%%%%%%%%%%%%%%%%%main body%%%%%%%%%%%%%%%%%%%%%%%%%%%%%%%%%%%%%%%%%%%%%%%%%%

\baselineskip16pt

The study of the transition regions separating adjacent quantum Hall (QH)
states is an active topic of research in the filed of two dimensional
electron systems\cite{Reviews1,Reviews2}. Several experimental groups have
studied the temperature dependence of the half width for the longitudinal
resistance $\rho _{xx}$, and of the maximum slope in the Hall resistance $%
\rho _{xy}$ between neighboring Hall plateaus\cite{Wei1,Koch1}. A remarkable
quantum phase transition has been demonstrated by showing a scaling behavior
in temperature: $\frac{d\rho _{xy}}{dB}\mid _{B_{c}}\propto T^{-\kappa }$
and $\Delta B\propto T^{\kappa }$ with critical exponent $\kappa $. However,
in sharp contrast to short range alloy potential scattering in InGaAs/InP
samples, AlGaAs-GaAs heterostructures has long range Coulomb scattering
which results in nonuniversality of the temperature exponent $\kappa $\cite%
{Koch2,Wei2}. Nevertheless, recently Li \textit{et al.} realized several Al$%
_{x}$Ga$_{1-x}$As-GaAs samples where dominant contribution of the disorder
is from short range random alloy potential fluctuation by varying Al
concentration $x$ and observed a perfect scaling with $\kappa =0.42$\cite{Li}%
.

There are extensive theoretical works to understand this quantum critical
phenomena and the actual value of $\kappa $. The easiest way to see this is
to consider a experiment where the magnetic field $B$ is being varied to
make the chemical potential move through $E_{c}$ to cause the
delocalization-localization transitions. In analogy with other localization
transitions, one expects a power-law divergence of localization length $\xi
\propto \left\vert B-B_{c}\right\vert ^{-v}$, where $v$ is the localization
critical exponent. Since the experimental data were measured at finite
temperature, the quantum phase coherence length sets an effective finite
system size $L$. Usually $L$ scales with temperature as $L\propto T^{-\frac{p%
}{2}}$ or $T^{-\frac{1}{z}}$, where $p$ or $z$ is called the dynamical
exponent. In the critical region, resistance tensor scales as $\rho
_{uv}=\rho _{c}f(L/\xi )$, where $f$ is a scaling function which can derived
from the microscopic calculation. Then one obtains $\rho _{uv}=\rho
_{c}f(\left\vert B-B_{c}\right\vert T^{-\kappa })$, the scaling function of
both the longitudinal resistance $\rho _{xx}$ and the Hall resistance $\rho
_{xy}$, where the exponent is expressed as $\kappa =p/2v$. Most of the
experimental work on the plateau transition in QH effect can be
quantitatively understood on the basis of the scaling law defined above,
which has sometimes been referred to as the one-parameter scaling theory\cite%
{Pruisken}.

Along this way, a great deal of theoretical work has been carried out to
study the quantity $v$\cite{Huckestein1,Bhatt}, and $p$\cite%
{Lee1,Lee2,Feng,Huckestein2}, using different models and different
calculation techniques. Several years ago, Sheng and Weng\cite{Sheng}
consider the Anderson Hamiltonian subjected to a strong magnetic field, use
Kubo formula, and find that the longitudinal resistivity in the critical
region well follows a simple exponential form:
\begin{equation}
\rho _{xx}=\rho _{c}\exp (-s),s=\left( \frac{c_{0}\left\vert
i-i_{c}\right\vert ^{v}}{T^{\frac{1}{z}}}\right) ,
\end{equation}%
where $\rho _{c}$ is the resistivity of critical point, $s$ is a scaling
variable, $z$ is the dynamical exponent, $v$ is the localization exponent, $%
i=hn/eB$ is the filling factor, is the critical point and $c_{0}$ is a
constant.

On the other hand, an alternative way to obtain this scaling phenomena is to
argue that the transport in the plateau transition is dominated by variable
range hopping (VRH) in the presence of Coulomb interaction. By assuming VRH
transport responsible for broadening of the $\sigma _{xx}$ peaks, Polyakov
and Shklovskii\cite{Polyakov} also arrives at an explicit expression of
scaling function%
\begin{equation}
\sigma _{xx}=\sigma _{0}\exp (-\sqrt{\frac{T_{0}}{T}})
\end{equation}%
with a temperature dependent prefactor $\sigma _{0}\propto 1/T$. Using an
appropriate definition of scaling variable, we obtain
\begin{equation}
\sigma _{xx}=\sigma ^{\ast }x\exp (-\sqrt{T^{\ast }x}),x=\frac{\left\vert
i-i_{c}\right\vert ^{v}}{T}
\end{equation}%
where $x$ is a scaling variable, $\sigma ^{\ast }$ and $T^{\ast }$ is a
constant. Why such hopping-based strong localization theory should have
validity in the scaling regime is not clear\cite{DasSarma}. However the
above two approaches looks very different, we are not aware of a direct
measurement to verify them in the same sample.

In a former paper, we have reported the plateau-to-plateau transitions and
corresponding exponents $\kappa $\cite{Tu1}. In the present paper, we focus
on the quantitative form of the scaling function $f$ in the transition
regime and compare it with the above two theoretical results. The sample we
measured was grown by molecular-beam epitaxy and consists of a $25$ nm wide
GaAs well bounded on each side by undoped and Si $\delta $-doped layers of Al%
$_{0.35}$Ga$_{0.65}$As. It has a low-temperature mobility $\mu =2.1\times
10^{5}$ cm$^{2}$/V s and the electron density $N_{s}$ is fixed at $2.8\times
10^{11}$ cm$^{-2}$. The sample is of high quality as it shows very strong
integer QH states. In order to measure $\rho _{xx}$ and $\rho _{xy}$, we use
a standard ac lock-in technique with electric current $I=10$ nA and
frequency of $13$ Hz. In the present experiment we use a $^{4}$He cryogenic
system from $1.7$K to $4$K, and further experiment in lower temperature
(several mK-$1$K region) using Oxford Dilution Refrigerator will appear soon%
\cite{Tu}. We would emphasize that although most of QH plateau
transition experiments are performed at lower temperature (below
$1$K), we are interested in the transition behavior at at somewhat
high temperature (above $1$K). Nevertheless, the sample can also
clearly show the scaling behavior in the plateau transitions at
temperature region above $1$K.

For simplicity, we focus on the transition from $i=3$ to $i=4$ integer QH
states. In Fig.(1) we plot $\rho _{xy}$ and $\rho _{xx}$ vs $B$ at different
temperature from $1.79$K to $2.69$K. A $\rho _{xx}$ peak that widens with $T$
and accompanying step in $\rho _{xy}$ clearly show the transition between
the neighboring plateaus around filling factor $i=3$, $4.$Next, we convert
the $\rho $'s to conductivity $\sigma $'s using the standard matrix
conversion,%
\begin{equation}
\sigma _{xx}=\frac{\rho _{xx}}{\rho _{xx}^{2}+\rho _{xy}^{2}},\sigma _{xy}=%
\frac{\rho _{xy}}{\rho _{xx}^{2}+\rho _{xy}^{2}}
\end{equation}%
and plot the $\sigma $ traces in Fig.(2).
\begin{figure}[tbp]
\includegraphics[height=7cm,width=1.0\columnwidth]{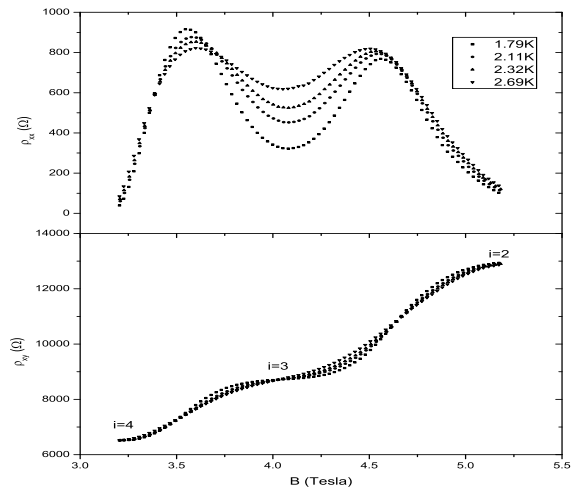}
\caption{The plateau to plateau transition in longitudinal
resistivity and Hall resistivity vs magnetic field.}
\end{figure}

\begin{figure}[tbp]
\includegraphics[height=7cm,width=1.0\columnwidth]{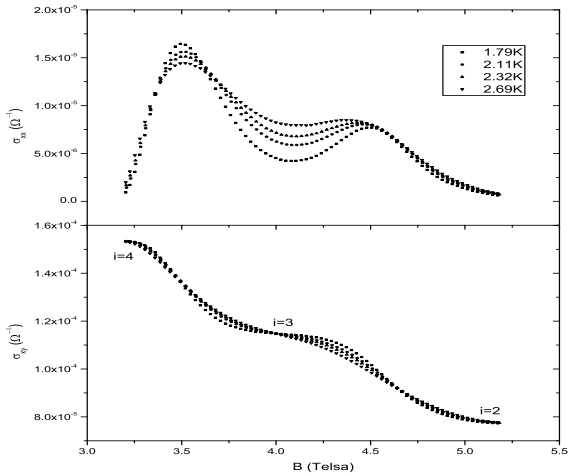}
\caption{Longitudinal conductivity and Hall conductivity vs magnetic field. }
\end{figure}

One way of extracting quantitative informations from the transition is by
conducting a scaling analysis of the data. In Fig.(3), $\rho _{xx}$ is
plotted as a function of scaling variable $s$ defined in Eq.(1). While the
critical point $i_{c}=hn/eB_{c}$ $\approx 3.4$ can be directly obtained form
the data, we vary the value of $z$ until we obtain the optimal collapse of
the $\rho _{xx}$ traces obtained at different $T$'s. We find that the
longitudinal resistance exhibits a exponential dependence such as Eq.(1) in
the critical region covering a range for $i_{c}<i<3.6$. The corresponding
scaling functions start to deviate from the exponential form beyond the
critical region. The resulting values of the critical exponent, $\kappa
=1/zv=0.69$, are indeed close to the values obtained from ordinary
approaches using the relations $\frac{d\rho _{xy}}{dB}\mid _{B_{c}}\propto
T^{-\kappa }$ and $\Delta B\propto T^{\kappa }$\cite{Tu1}.
\begin{figure}[tbp]
\includegraphics[height=7cm,width=1.0\columnwidth]{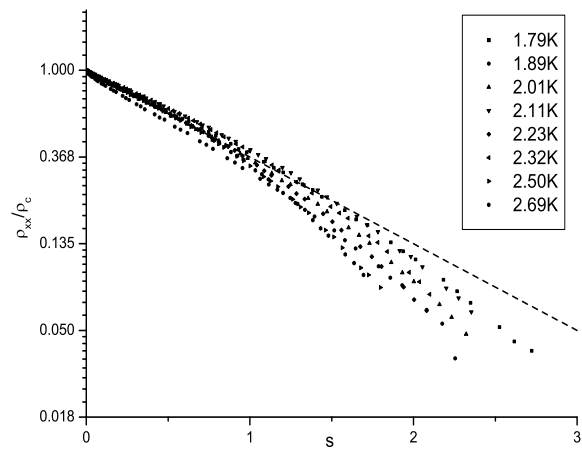}
\caption{$\protect\rho _{xx}$ as a function of the scaling variable $s$. We
use the data of 3-4 transition. Dashed line is the function of Eq.(1) of
variable $s$. }
\end{figure}

In Fig.(4), $\sigma _{xx}$ is plotted as a function of scaling variable $x$
defined in Eq.(2). We vary the value of $v$ until all experiment points of $%
\sigma _{xx}$ fall onto a straight line as represented by the
scaling function in Eq.(3). We find that the longitudinal
conductivity $\sigma _{xx}$ exhibits a scaling function dependence
such as Eq.(3) in the transition region covering a wide range from
$3.5<i<3.8$. The corresponding scaling functions start to deviate
from the form of Eq.(3) beyond the above region. The resulting
values of the critical exponent, $v=2.35$, are indeed close to the
values obtained from Koch. \textit{et al}\cite{Koch2,Haug2}.
\begin{figure}[tbp]
\includegraphics[height=7cm,width=1.0\columnwidth]{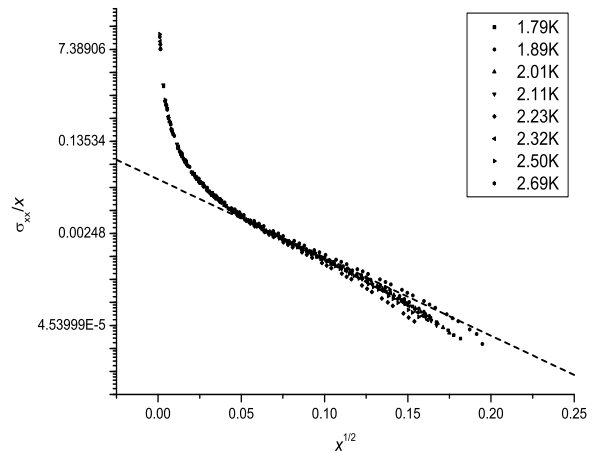}
\caption{$\protect\sigma _{xx}$ as a function of the scaling variable $x$.
We use the data of 3-4 transition. Dashed line is the function of Eq.(3) of
variable $x$. }
\end{figure}

In our experiment, we obtain $\rho _{xx}$ or $\sigma _{xx}$ data to
study the plateau transition in the region $3\leq i\leq 4$. From the
above analysis, we find that both two scaling form are valid for
different regions of plateau transition (one is near the critical
point $i_{c}$ $\approx 3.4$, $3.4<i<3.6$, the other is slightly far
from the critical point $3.5<i<3.8$). Here we proposed an possible
physical explanation to support this difference. In integer QH
effect, since the Fermi energy is in a region of localized states
when the Hall resistivity is quantized and the longitudinal
resistivity vanishes, it can be concluded from the wide plateau that
most of the electron states are localized at low temperatures.
Actually Paalanen \textit{et al.} estimated that, in an AlGaAs-GaAs
heterostructres at $50$mK, 95\% of the states in each Landau level
are localized\cite{Paalanen}. The mechanism for the conductivity is
variable range hopping (VRH) in the presence of Coulomb interaction.
Thus we may expect that VRH can dominates the conductivity in the
transition regime, when the localization length $\xi $ becomes much
smaller than the effective temperature length $L$. This means
Eq.(3) is valid in the region which is slightly far from the critical point $%
i_{c}$ (eg. $\,3.5<i<3.8$ in our sample). On the other hand,
metal-insulator transition dominates the conductivity in the
transition regime, when the localization length $\xi $ becomes much
large than the effective temperature length $L$. Then Eq.(1) is
valid in the region which is near the critical point $i_{c}$ (eg.
$\,i_{c}<i<3.6$ in our sample). Thus the two different scaling
function are valid in different regions in the plateau transition.

In conclusion, we have measured the temperature dependence of the
longitudinal resistivity $\rho _{xx}$ of a two-dimensional electron system
in the regime of the quantum Hall plateau transition. We extracted the
quantitative form of scaling function for $\rho _{xx}$ and compared it with
two alternative theoretically proposed scaling functions. Further we
determine different regions where the two alternative theoretically proposed
scaling functions are valid in the same sample. However we focus on the
behavior at high temperatures, similar experiment in lower temperature
(several mK-$1$K region) would be discussed in the near future\cite{Tu}.

We thank Prof. H. W. Jiang (UCLA), Prof. R. R. Du (Rice), Prof. D.
N. Sheng (CSU), and Dr. Y. Chen (Princeton) for helpful discussions.
This work is supported by National Fundamental Research Program, the
Innovation funds from Chinese Academy of Sciences, National Natural
Science Foundation of China (No.60121503 and No.10604052).


\begin{thebibliography}{99}
\bibitem{Reviews1} S. L. Sondhi \textit{et al.}, Rev. Mod. Phys. \textbf{69}%
, 315 (1997).

\bibitem{Reviews2} B. Huckestein, Rev. Mod. Phys. \textbf{67}, 357 (1995).

\bibitem{Wei1} H. P. Wei \textit{et al.}, Phys. Rev. Lett. \textbf{61}, 1294
(1988).

\bibitem{Koch1} S. Koch \textit{et al.}, Phys. Rev. B \textbf{43}, 6828
(1991).

\bibitem{Koch2} S. Koch \textit{et al.}, Phys. Rev. Lett. \textbf{67}, 882
(1991).

\bibitem{Wei2} H. P. Wei \textit{et al.}, Phys. Rev. B \textbf{45}, 3926
(1992).

\bibitem{Li} W. L. Li \textit{et al.}, Phys. Rev. Lett. \textbf{94}, 206807
(2005).

\bibitem{Pruisken} A. M. M. Pruisken, Phys. Rev. Lett\textit{.} \textbf{61,}
1297 (1988).

\bibitem{Huckestein1} B. Huckestein and B. Kramer, Phys. Rev. Lett. \textbf{%
64}, 1437 (1990).

\bibitem{Bhatt} Y. Huo ang R. N. Bhatt, Phys. Rev. Lett. \textbf{68}, 1375
(1992).

\bibitem{Lee1} D. H. Lee \textit{et al.}, Phys. Rev. Lett. \textbf{70}, 4130
(1993).

\bibitem{Lee2} D. H. Lee and Z. Q. Wang, Phys. Rev. Lett. \textbf{70}, 4110
(1993).

\bibitem{Feng} H. L. Zhao and S. Feng, Phys. Rev. Lett. \textbf{70}, 4134
(1993).

\bibitem{Huckestein2} B. Huckestein and M. Backhus, Phys. Rev. Lett. \textbf{%
82}, 5100 (1999).

\bibitem{Sheng} D. N. Sheng and Z. Y. Weng, Phys. Rev. Lett. \textbf{83,}
144 (1999).

\bibitem{Polyakov} D. G. Polyakov and B. I. Shklovskii, Phys. Rev. Lett.
\textbf{70}, 3796 (1993).

\bibitem{DasSarma} S. Das Sarma, in \textit{Perspectives in Quantum Hall
Effects}, edited by S. Das Sarma and A. Pinzuk (Wiley, New York, !997).

\bibitem{Tu1} T. Tu \textit{et al.}, submitted (2007).

\bibitem{Haug2} F. Hohls, \textit{et al.}, Phys. Rev. Lett. \textbf{88},
036802 (2002).

\bibitem{Paalanen} M. A. Paalanen \textit{et al.}, Phys. Rev. B \textbf{25},
5566 (1982).

\bibitem{Tu} T. Tu \textit{et al.}, in prepare (2007).
\end{thebibliography}
\end{document}